\documentclass{article}

\usepackage{PRIMEarxiv}

\usepackage[utf8]{inputenc} 
\usepackage[T1]{fontenc}    
\usepackage{hyperref}       
\usepackage{url}            
\usepackage{booktabs}       
\usepackage{amsfonts}       
\usepackage{nicefrac}       
\usepackage{microtype}      
\usepackage{lipsum}
\usepackage{fancyhdr}       
\usepackage{graphicx}       
\graphicspath{{media/}}     

\usepackage{booktabs}
\usepackage{comment}

\title{Who Owns the Output? Bridging Law and Technology in LLMs Attribution} 

\author{
  Emanuele Mezzi \\
  Ethikon Institute \\
  mezziemanuele@outlook.com
  \And
  Asimina Mertzani \\
  Ethikon Institute \\
  mertzanisemina@gmail.com
  \And
  Michael P. Manis \\
  Ethikon Institute \\
  michailmanis@gmail.com
  \And 
  Siyanna Lilova \\
  Ethikon Institute \\
  lilova.siyanna@gmail.com
  \And
  Nicholas Vadivoulis \\
  Ethikon Institute \\
  nikvadiv@gmail.com
  \And
  Stamatis Gatirdakis \\
  Ethikon Institute \\
  gatirdakis.s@gmail.com \\
  \And
  Styliani Roussou \\ 
  Ethikon Institute \\ 
  stellaroussou21@hotmail.com
  \And
  Rodayna Hmede \\
  Ethikon Institute \\
  rodaynah.hmede@gmail.com
}

\begin{document}
\maketitle

\begin{abstract}
Since the introduction of ChatGPT in 2022, Large language models (LLMs) and Large Multimodal Models (LMM) have transformed content creation, enabling the generation of human-quality content, spanning every medium, text, images, videos, and audio. The chances offered by generative AI models are endless and are drastically reducing the time required to generate content and usually raising the quality of the generation. However, considering the complexity and the difficult traceability of the generated content, the use of these tools provides challenges in attributing AI-generated content. The difficult attribution resides for a variety of reasons, starting from the lack of a systematic fingerprinting of the generated content and ending with the enormous amount of data on which LLMs and LMM are trained, which makes it difficult to connect generated content to the training data. This scenario is raising concerns about intellectual property and ethical responsibilities. To address these concerns, in this paper, we bridge the technological, ethical, and legislative aspects, by proposing a review of the legislative and technological instruments today available and proposing a legal framework to ensure accountability. In the end, we propose three use cases of how these can be combined to guarantee that attribution is respected. However, even though the techniques available today can guarantee a greater attribution to a greater extent, strong limitations still apply, that can be solved uniquely by the development of new attribution techniques, to be applied to LLMs and LMMs. 
\end{abstract}


\section{Introduction}
\label{sec:introduction}

Large Language Models (LLMs) \cite{minaee2024largelanguagemodelssurvey} and Large Multimodal Models (LMMs) \cite{wang2024comprehensive} significantly transformed the landscape of content creation, leading to the automation of complex tasks, drastically reducing the effort and the amount of time necessary to perform them. These models which rely on an enormous amount of data for their training, exhibit an unprecedented ability to generate human-quality content, making them powerful tools across industries, including media, law, and education. 

At the same time, their use, raises critical questions about the origins of AI-generated content, as the immense scale and opacity of the training data complicate the task of discerning the sources that contribute to generated outputs \cite{zhao2023survey}.

Connecting the AI-generated content to both the individuals responsible for its generation and to the training data that allows for its generation is called \textit{attribution}. Attribution is crucial for assuring intellectual property, accountability, and transparency, which are given by determining who instructed the model to generate specific content and which training data allowed the model to generate specific content. Although several techniques have been integrated to reach attribution such as digital watermarking \cite{wadhera2022comprehensive}, AI model fingerprinting \cite{xu2024instructional}, Blockchain-Based Attribution \cite{oham2018blockchain}, explainable AI \cite{zhao2024explainability}, techniques that guarantee deterministic complete attribution are still to be achieved. By aligning these technical methods with clear legal frameworks and ethical guidelines,
stakeholders would enhance accountability in AI systems, ensuring that attribution mechanisms are effective, transparent, and respected across various jurisdictions. To effectively bridge the legal and technical gaps in AI attribution, several \textit{key policy recommendations} should be implemented. The \textit{interaction between legal frameworks and technical standards must be deeply embedded within both fields through collaborative workshops and interdisciplinary research initiatives}. Then, \textit{educational curricula should be revamped to provide engineers and legal professionals with advanced training in AI technology and ethics}. 

Additionally, considering temporary protection mechanisms, such as \textit{provisional patents}, can facilitate technology adoption while balancing intellectual property concerns. \textit{Promoting empirical research on the alignment of local regulations with national legal frameworks} will aid in developing adaptable policies. \textit{Establishing clear regulatory guidelines for AI development, adopting ethical standards, and enhancing public awareness through stakeholder engagement are also crucial steps.} Together, these recommendations foster collaboration, promote innovation, and ensure policies remain relevant in the rapidly evolving AI landscape. 

Attribution to AI-generated content is a pressing issue that intersects technical, ethical, and legal dimensions. As AI systems produce content derived from vast and often opaque datasets, tracking the origin and authorship becomes challenging, leading to potential concerns over copyright infringement, loss of recognition for original creators, and accountability. Ensuring reliable attribution is crucial not only for protecting intellectual property rights but also for establishing trust in AI technologies and fostering responsible use of generative AI.

This paper aims to address the complexities of attribution in AI-generated content by clarifying how attribution mechanisms work, proposing a legal framework to support fair use and accountability, and benchmarking the effectiveness of various technical methods for attribution. Through this interdisciplinary approach, the paper seeks to advance the dialogue on AI ethics, intellectual property, and the integration of legal and technical solutions to ensure a robust and enforceable attribution system. 

\section{Legal Background}
\label{sec:legal-background}
\subsection{International Copyright Law and AI}
\label{subsec:international-copyright-law-and-AI}
The challenges AI poses to copyright law are multidimensional and further complicated by the territorial nature of copyright law, which means the law is enacted and enforced through national laws. Nonetheless, the international copyright legal framework is the cornerstone for the current debates and interpretations of the relationship between AI and copyright law. International copyright treaties such as the Berne Convention for the Protection of Literary and Artistic Works (Berne Convention), the Agreement on Trade-related Aspects of Intellectual Property Rights (TRIPS), and the WIPO Copyright Treaty (WCT), are relevant to the legal regime of AI outputs, the legal use of copyrighted AI inputs, as well as the right to attribution in the context of AI training. 

\subsection{Authorship, attribution, and the Berne Convention} 
\label{subsec:authorship-attribution-and-Berne Convention}
\begin{figure}
    \centering
    \includegraphics[width=0.80\linewidth]{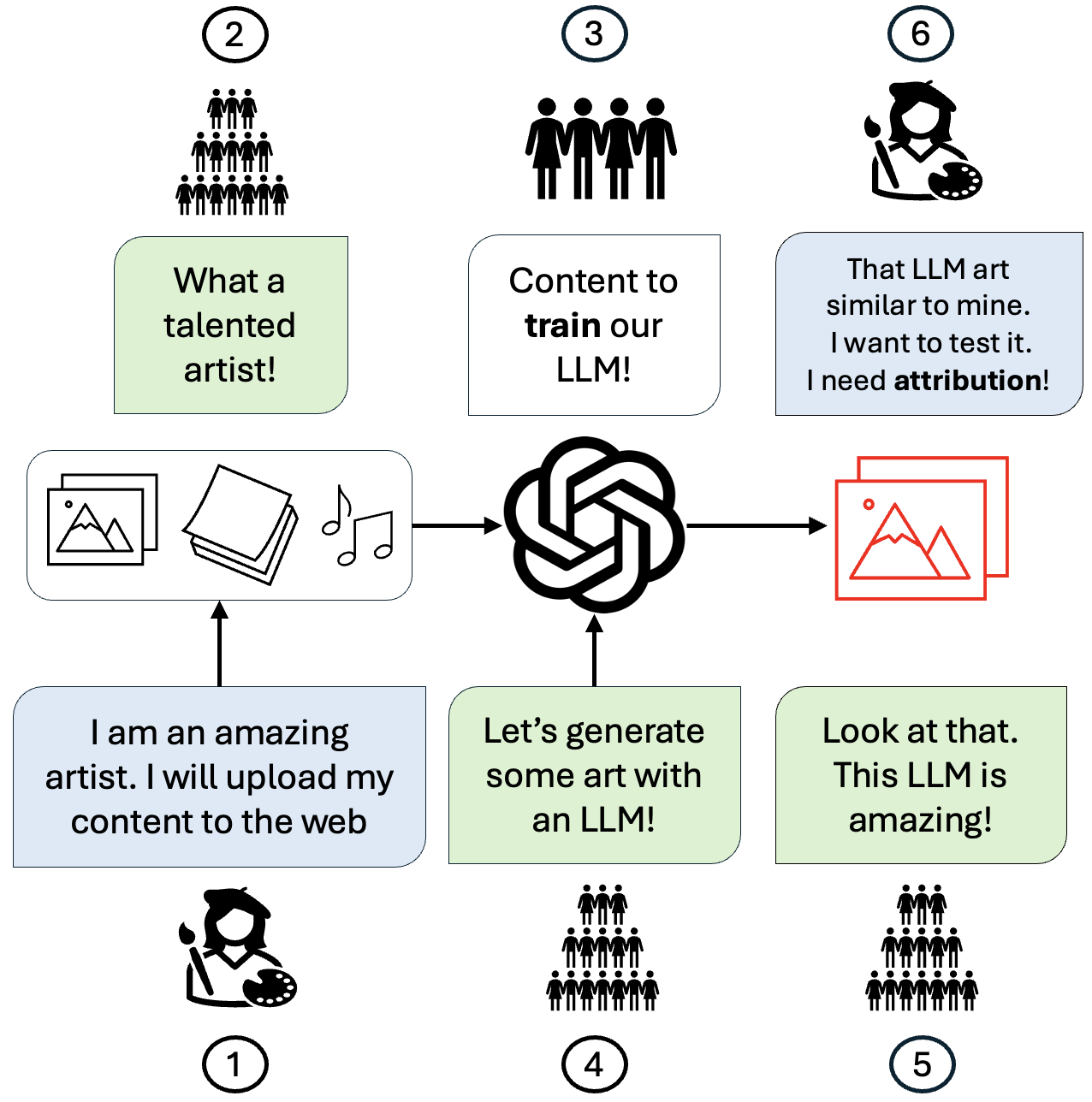}
    \caption{In this figure we show how attribution can help protecting the copyright and the work of content creators.}
    \label{fig:attribution-motivation}
\end{figure}

When considering AI outputs and copyright law, a main subject of dispute is whether such outputs merit copyright protection, especially in cases where no human creativity is involved (Figure \ref{fig:attribution-motivation}) (i.e., entirely AI-generated content) and if so - who should merit such protection (Section \ref{subsec:international-copyright-law-and-AI}). On the subject of non-human authorship and international copyright law, already in 1991, professor Sam Ricketson clarified that “the need for the author[s] to be a human being and for there to be some intellectual contribution” is a longstanding assumption in national copyright laws \cite{ricketson19911992}. This position has been echoed by many scholars who reject the concept of “machines as authors”. According to Ricketson, even though there is no definition of ‘author’ in the Berne Convention, the human authorship requirement is so foundational for the international copyright system that without it copyright would become devoid of its ‘soul’ \cite{ricketson19911992}.

Not only human authorship but also attributing the author as such lies at the heart of copyright law. As Gervais points out, a crucial first step in the evolution of the concept of ‘authorship’ was the ability to acknowledge the author by the name \cite{gervais2019machine}. In other words, naming the author, or attribution, was crucial for the emergence of the idea of authorship, and therefore of copyright itself, establishing deep connections between the author and their work. This highlights not only the issue of attributing works to an AI system or an LLM but also the fundamental importance of the right to attribution, called by Ginsberg ‘the most moral of rights’ \cite{ginsburg2016most}.

The right to attribution is regulated in Article 6bis of the Berne Convention: “[i]ndependently of the author's economic rights, and even after the transfer of the said rights, the author shall have the right to claim authorship of the work...”. 
According to Article 6bis, the duration of the attribution right should be at least the same as the term of economic rights but all other specifics as to the scope and application of the right are left to national legislation \cite{ricketson19911992}. 

In some countries, like Bulgaria for example, the right to attribution has an exceptionally wide scope in comparison to other economic and moral rights of authors, being with an indefinite scope and applying to any use of the work \cite{sarakinov1994new}. In addition, it is common that the right to attribution extends to adaptations based (substantially) on the original work. This is especially relevant to the discussion on the nature of AI output and whether it can be considered a derivative work based on the original input.

\subsection{Copyright exceptions, three-step test, and TRIPS} 
While the Berne Convention introduced the so-called ‘three-step test’ for copyright exceptions and limitations in the context of the right to reproduction, the TRIPS Agreement extended it to other exclusive rights \cite{lilova2021copyright}.

The three-step test under TRIPS is key to understanding the foundational principles that allow the legal use of copyrighted content without the author’s permission, including for training AI, under statutory exceptions and limitations or the fair use doctrine. Any (new) exception for the use of copyrighted content as AI input, whether on an international or national level, should therefore comply with the three-step test \cite{lilova2021copyright}. The three-step test is particularly important not only because it impacts the drafting and implementation of exceptions and limitations but also because it serves as an important interpretation tool for domestic courts \cite{rosati2023no}. 

According to the three-step test, Member States may introduce copyright exceptions only “in certain special cases which do not conflict with a normal exploitation of the work or other subject matter and do not unreasonably prejudice the legitimate interests of the right holder.” This has been interpreted to mean that: (1) exceptions may not be overly broad, thus applicable only in “certain special cases”; (2) exceptions may not “rob right holders of a real or potential source of income that is substantive” which would “conflict with normal exploitation of the work”; and (3) exceptions may not “do disproportional harm to the rights holders,” hence may not “prejudice legitimate interests” of the right holder \cite{hugenholtz2012conceiving}.

Importantly, this suggests that the TDM exception in Europe and the fair use doctrine interpretation in US courts cannot justify reproductions that lead to applications that substitute or otherwise significantly economically compete with the protected material used for AI training (see Section \ref{subsubsec:use-of-copyrighted-works-AI-EU}). Even though it can be argued that this is exactly what is currently happening with the widespread use of AI-generated content \cite{novelli2024generative}.  It should be noted, however, that TRIPS does not cover the right to attribution, as Article 6bis of the Berne Convention is specifically excluded from the scope of the agreement. This means that Member States are free to either include stricter restrictions for moral rights or extend the three-step test rationale to all authors’ rights, further complicating the international regime applicable to exceptions of the attribution right.

\subsection{Electronic rights management, digital attribution, and WCT} 
The WCT and WIPO Performances and Phonograms Treaty (WPPT), termed together as the WIPO ‘internet treaties’ include rules to adapt international copyright law to the digital environment. One of these rules is the obligation for Member States to provide “adequate and effective remedies against any person who knowingly” removes or alters “electronic rights management information” without authority, or distributes, imports for distribution, broadcasts or communicates to the public works or copies of works without authorization, knowing that electronic rights management information has been removed or altered without authority. This means that the removal of rights management information, which often occurs during the training stage of an AI model, could constitute a separate violation of a copyright holder's rights \cite{gervais2024heart}. 

According to the treaties, ‘rights management information’ means “information which identifies the work, the author of the work, the owner of any right in the work, or information about the terms and conditions of use of the work, and any numbers or codes that represent such information, when any of these items of information is attached to a copy of a work...” The removal of authorship data in AI inputs can therefore constitute a separate infringement under the national legislation transposing the WCT and WPPT provisions. This can provide authors with an alternative path for protecting their right to attribution in cases where national law does not offer protection for moral rights, or where such protection is limited. This is already seen in the US, where there is no full statutory protection of the right to attribution, and plaintiffs in cases against LLM providers are relying on the removal/modification of copyright management information as a way to enforce their right to attribution. The US Copyright Office has even recognized that the removal of digital rights management information is a practical stand-in for attribution in the technological context (see Section \ref{subsubsec:fair-use-ai-gen-cont}). However, to prove infringement, the plaintiff must show that the defendant knew or had reasonable grounds to know that the removal would “induce, enable, facilitate or conceal an infringement” of a copyright. It therefore remains to be seen how effective these provisions will be in enforcing attribution in the context of AI model training both in the US and internationally.

\subsection{Legal status of AI-generated works} 
\label{subsec:legal-status-ai-gen-works}
\subsubsection{Authorship \& Artificial Intelligence: The Role of Non-Human Creators in Copyright Law from an EU and US Perspective}
The discussion around authorship in the context of artificial intelligence (AI) has become a crucial point of legal disputes, particularly as AI-generated and mostly AI-assisted outputs become increasingly dominant. As the creative capabilities of AI systems continue to evolve, the challenge of determining the rightful owner of the copyright attached to such works intensifies. Specifically, \textit{the question of whether non-human entities (particularly AI systems) can be recognized as authors under copyright law is of paramount importance}. This section will explore how courts and regulatory bodies have approached the concept of non-human authorship and discuss its implications for AI-generated works within the framework of the European Union (EU) and the United States (US).

\subsubsection{The EU Approach}
In recent years, various authorities and organizations have begun issuing guidelines to address the challenges posed by AI-generated content. The European Commission has acknowledged the necessity for a comprehensive approach to intellectual property in the digital age, proposing regulatory frameworks that consider the unique attributes of AI technologies. These guidelines aim to clarify the role of AI in the creative process and provide a roadmap for determining authorship in scenarios where human involvement is limited or does not exist. The emergence of collaborative platforms enabling users to interact with AI tools necessitates a reevaluation of authorship. In scenarios where a user initiates a creative process using an AI system, a critical question arises: who should be considered the author? The distinction between active and passive involvement becomes paramount. If a user merely inputs prompts without exercising substantial creative judgement, their claim to authorship may be significantly weakened. Conversely, if the user’s decisions contribute meaningfully to the final output, they may be recognized as co-authors alongside the AI developer.

The EU perspective has recently been significantly shaped by the AI Act \cite{madiega2021artificial}, which introduces a regulatory framework specifically designed for AI technologies. This Act emphasises that while AI can assist in the creative process, the authorship of any output remains inherently tied to human creators. This viewpoint is further supported by Article 2 of Directive 2009/24/EC, which defines authorship in relation to natural persons and explicitly excludes non-human entities from holding copyright. The AI Act establishes a Code of Practice that aims to clarify the role of AI in creative processes, addressing the challenges associated with determining authorship when AI systems generate content with limited human input. Consequently, it is imperative to evaluate the extent of human involvement in the creative process; only significant creative contributions may justify recognition as a co-author alongside the AI, thereby maintaining a clear distinction between human creativity and automated outputs. In the judicial arena, the European Court of Justice (ECJ) has reinforced the principle that authorship is inherently linked to human creativity. In the case of Martin Luksan v Petrus van der Let, the ECJ addressed the nuances of authorship and originality in the context of database rights, emphasizing that the notion of an "author" necessitates a human creator. This decision resonates in subsequent discussions on AI-generated works, highlighting the court's reluctance to extend authorship status to non-human entities. The ruling underscores the understanding that legal recognition of authorship requires a degree of human intellectual contribution that is often absent in purely AI-generated outputs. Similarly, the Hewlett-Packard Belgium SPRL v Reprobel SCRL case underscored the importance of human intervention in the creative process. The court ruled that reproduction rights are inherently tied to human authorship, thereby establishing a legal precedent that AI, in its current form, cannot claim authorship or rights under existing copyright laws. This perspective is crucial for interpreting the role of AI as a tool rather than as a creator in its own right.

In conclusion, as AI technologies continue to evolve and permeate various creative industries, the interpretation of authorship in the context of EU copyright law remains a dynamic and evolving area. The principles established in cases such as Martin Luksan v Petrus van der Let and Hewlett-Packard Belgium SPRL v Reprobel SCRL, alongside the guidelines provided by Directive 2009/24/EC and the emerging regulations under the AI Act, highlight the essential requirement of human creativity in the attribution of authorship. Moving forward, the challenge lies in balancing the legal frameworks with the realities of AI involvement in creative processes, ensuring that both human creators and AI developers are recognized appropriately while fostering innovation in the digital landscape. This exploration enhances the pressing need for a comprehensive legal approach that can adapt to the complexities introduced by AI in the creative domain, paving the way for fair and equitable attribution of authorship.

\subsubsection{The US Approach}
The U.S. Constitution explicitly grants Congress the power to secure, for a limited time, the exclusive rights of authors to their writings. This provision is the foundation for the Copyright Act, which aims to protect “original works of authorship.” While the Constitution and the Copyright Act do not clearly define who can be considered an "author," the U.S. Copyright Office has established that copyright can only be granted for works created by a human being.

This interpretation has been reinforced by various court rulings. For instance, in the Naruto case \cite{hooker2020naruto}, a monkey that took a series of photographs was found to lack the legal standing to file a lawsuit under the Copyright Act, illustrating the current limitation that nonhuman entities cannot claim authorship. Similarly, in the Urantia Foundation case involving a book supposedly inspired by celestial beings, the court ruled that some degree of human creativity must be present for a work to be eligible for copyright protection \cite{rhee1998urantia}. Furthermore, a living garden was determined to be ineligible for copyright, as it did not have a human author associated with its creation. These cases collectively underscore the prevailing legal standpoint that only human authors can hold copyright over original works.

Given that, it is possible for works produced by individuals utilizing generative AI to be eligible for copyright protection. However, this eligibility is contingent upon the extent of human involvement in the creative process. Recent legal cases and updated guidance from the Copyright Office suggest a growing scepticism regarding the recognition of human authorship in instances where an AI program generates material based solely on text prompts. In September 2022, artist Kris Kashtanova registered a copyright for her graphic novel, which included images generated in response to her prompts submitted to the AI software Midjourney. A month later, the Copyright Office initiated cancellation proceedings, noting that Kashtanova had not disclosed her use of AI technology in her registration application. In her defence, she argued that the images were created through a "creative, iterative process," indicating a significant level of human creativity and decision-making. On February 21, 2023, the Copyright Office reached its decision, concluding that the images were not eligible for copyright protection. They asserted that Midjourney, the AI program, was the actual author of the "visual material," rather than Kashtanova. Following this, in March 2023, the Copyright Office released new guidance, stating that when AI determines the expressive elements of its output, the resulting material cannot be considered the product of human authorship. This guidance represents a significant clarification in the ongoing discussion regarding copyright and the role of artificial intelligence in creative processes.

In general, AI systems undergo a training process that enables them to generate various forms of artistic expression, including literature and visual art. This process involves feeding the AI large datasets comprised of text, images, and other content acquired from the vast resources available on the internet. During training, the AI program creates digital copies of existing works, a necessary step for analyzing the vast amount of information it receives. The U.S. Patent and Trademark Office has noted that this training inherently involves reproducing either complete works or significant portions of those works. For instance, OpenAI has publicly acknowledged that its programs are trained on extensive datasets that include copyrighted materials. Specifically, they emphasize that the training process begins with making copies of the data intended for analysis. While OpenAI has recently introduced an option to exclude certain images from being used in the training of future image generation models, it's important to recognize that the act of creating these copies without obtaining permission may violate the exclusive rights of copyright holders to reproduce and distribute their creative works. This raises important questions about copyright and intellectual property in the context of advancing AI technology.

The critical question surrounding the use of generative AI is who should be held accountable if its outputs violate the copyrights of existing works. Current legal frameworks suggest that both the individual using the AI and the company that developed the technology could be deemed responsible for such infringements.

For instance, if a user is found to be directly liable for copyright infringement by generating content that illegally replicates protected material, the AI company might also be implicated. This potential liability arises from the legal principle known as "vicarious infringement." Under this doctrine, a party can be held liable if they have the authority and capability to oversee the infringing actions and if they also have a direct financial stake in the activities occurring through their technology. This complicates the landscape of accountability, as it raises important questions about the responsibilities of both users and developers in the realm of AI-generated content.

\subsubsection{AI-generated content as a derivative work}
The rapid development of artificial intelligence (AI) tools capable of creating complex content, from images and text to music and multimedia, has spurred significant debate over the legal and ethical status of AI-generated works. Central to this discussion is the question of whether AI outputs should be considered “derivative works” under copyright law, as they often rely on vast amounts of preexisting, human-created content to generate something new. This reliance raises questions about the potential for copyright infringement, especially as AI tools become more sophisticated in their ability to replicate particular styles, structures, or elements drawn from the training data used in their development.
Under copyright law, derivative works are those that are based on existing content but altered enough to provide a new expression, form, or purpose distinct from the original. A traditional example might include a film adaptation of a novel or a reinterpretation of a song. For a work to be deemed derivative, it typically must contain identifiable elements of the original while adding a unique contribution. With AI, however, determining this distinct contribution is complicated by the absence of human authorship. AI systems are built to analyse patterns and recombine elements from large datasets, which often include copyrighted materials. This means that, although AI outputs may look or sound original, they are often directly shaped by the existing content fed into the system.

One of the primary legal questions involves whether AI-generated content transforms the original work enough to qualify as “new” or if it instead reproduces protected elements in a way that infringes copyright \cite{creata2019directive}. Courts \cite{oratz2024initial} typically analyse the amount of original material used, the intent and market impact of the use, and whether the final output offers new expression or value. In some jurisdictions, such as the United States, the doctrine of “transformative use” under fair use guidelines provides a pathway for determining whether a work is sufficiently altered from the original. This allows certain uses of copyrighted materials if they add enough new expression or purpose, although the extent of such protection for AI-generated content remains an open question.

In addition to transformative use, AI-generated content raises broader concerns over the potential devaluation of human creativity. Critics argue that, by relying on massive datasets of existing works, AI-produced content may saturate markets and diminish the unique contributions of individual artists or creators. This has led to concerns that AI could become a substitute for human talent, especially as it grows capable of replicating specific artistic techniques or styles. While some argue that AI tools are simply a modern form of creative expression, others caution that unchecked reliance on copyrighted datasets risks eroding the incentive for original human creation.
There is also growing legal scrutiny over how AI training data is gathered and used. In the European Union, for instance, the Digital Single Market Directive grants copyright holders the option to prevent their works from being used in “text and data mining” (TDM) by AI systems without authorization. This regulation reflects the sensitivity surrounding the reuse of copyrighted content in commercial AI models and highlights the need for a balanced approach that respects intellectual property while fostering innovation. In other jurisdictions, including the United Kingdom, TDM is allowed primarily for non-commercial research purposes, leaving commercial applications without clear legal protections unless proper licensing agreements are in place.

As AI models continue to advance, legislative bodies and courts face the challenge of redefining copyright protections in light of these new technological capabilities. The aim is to strike a balance that enables technological progress without diminishing the value of human-created works. Going forward, it will be necessary for legal frameworks to address the extent to which AI-generated works rely on existing data and whether such reliance classifies these outputs as derivative or transformative. In the meantime, copyright law’s existing principles of fair use, transformative use, and derivative work classifications remain key in guiding how AI-generated content is interpreted and protected.

\subsection{Use of Copyrighted Content for AI Input and Attribution}
\subsubsection{Use of copyrighted works as AI input in the EU} 
\label{subsubsec:use-of-copyrighted-works-AI-EU}
The EU's Artificial Intelligence Act (AI Act) \cite{smuharegulation} establishes a governance framework for data used in the training, validation, and testing of high-risk AI systems, recognizing that input data is crucial for the accuracy and functionality of AI models. This reliance on extensive datasets raises substantial copyright issues, as much of the content required for training may be protected by copyright or related rights, such as the sui generis database right. While the AI Act sets out data governance standards and includes some transparency and copyright-related rules, it is the EU’s copyright aquis that principally regulates the use of artistic content for AI training.  

Under the current EU copyright framework, there are several options for works to be legally used as input for AI training. First, developers can use public domain works, including materials whose copyright term has expired or content inherently unprotected by copyright, such as facts, data, and legislative texts. Alternatively, open-source licenses, particularly those permitting commercial use, such as the Creative Commons Attribution (CC-BY) license, offer a structured means of access to content, typically requiring only that proper attribution be provided. For materials under copyright, individual licensing or collective rights management schemes remain the main options for lawful use of copyrighted content unless there is an applicable copyright exception or limitation. The EU’s Digital Single Market (DSM) Directive \cite{creata2019directive} introduced two such exceptions for the use of copyrighted works for text and data mining (TDM). 

\subsubsection{Scope and applicability of the TDM exceptions} 
The DSM Directive provides two exceptions for TDM that enable the reproduction and extraction of lawfully accessible copyrighted content without prior consent from the rights holder. Despite some initial uncertainty about whether the TDM exceptions apply to AI training, the AI Act explicitly extends them to general-purpose AI models, requiring all providers bringing such models to the EU market to adhere to these provisions, regardless of where copyright-related training activities occur. Article 2(2) of the DSM Directive broadly defines TDM as “any automated analytical technique aimed at analysing text and data in digital form to generate information which includes but is not limited to patterns, trends and correlations.” This definition covers the key stages of TDM: (1) accessing data, (2) extracting or copying content, and (3) mining of text and/or data and knowledge discovery \cite{rosati2018exception}. TDM techniques can, therefore, infringe both the reproduction and extraction rights of the author and potentially also their adaptation rights as content is usually reformatted to be made suitable for AI training. 

Both Articles 3 and 4 of the DSM Directive encompass the acts of extraction and reproduction. Article 3 DSM grants an exception for scientific research by research organizations and cultural institutions, which cannot be restricted by contract. Article 4, however, allows a broader exception for TDM with no restriction on beneficiaries or purposes. Although it permits rights holders to reserve their rights through an opt-out mechanism. Art. 4(3) applies only on condition that right holders have not expressly reserved their rights “in an appropriate manner, such as machine-readable means in the case of content made publicly available online”. This is done in practice usually by adding a robot.txt metadata to website content but ambiguity still remains regarding what constitutes an effective opt-out \cite{keller2023generative}. Efforts to standardize opt-out procedures, such as the “TDM Reservation Protocol” seek to address this gap. The Regional Court of Hamburg in Kneschke v. LAION clarified (although in non-binding obiter dicta) that a natural language opt-out in a website’s terms of use should also be sufficient to exclude the commercial TDM exception.

There is still little Member State case law concerning the TDM exceptions, leaving many questions about their applicability in practice open. This includes also how courts will interpret the general restriction under Article 5(5) InfoSoc Directive, which applies to the two TDM exceptions via a reference in Article 7(2) DSM Directive. Article 5(5) InfoSoc Directive, reiterates the three-step test applicable in EU copyright law, stating that exceptions and limitations can “only be applied in certain special cases which do not conflict with a normal exploitation of the work or other subject-matter and do not unreasonably prejudice the legitimate interests of the rightholder.” Recent CJEU case law imposes an obligation upon national courts to consider the three-step test when deciding on the applicability of exceptions and limitations, evaluating whether the acts of the defendant satisfy the requirements of Article 5(5). In other words, it should be evaluated whether the use of protected material for AI training prejudices the legitimate interests of rightsholders (including economic ones), which is arguably often the case with generative AI. 

\subsubsection{The right to attribution for AI inputs} 
Attribution has emerged as an important issue in the legal discourse surrounding generative AI, driven by creators' demands to be credited for their contributions when their works are used as input for AI model training \cite{kahveci2023attribution}. This concern is amplified by several US cases related to so-called ‘output leakage’, where AI outputs directly replicate original content with identifiable elements like watermarks. In the Getty Images case, AI-generated images replicated the Getty watermark, revealing that the AI model had trained on Getty’s copyrighted database without permission or attribution. Similarly, GitHub’s CoPilot has faced scrutiny for generating code snippets identical to copyrighted code from public repositories, raising questions about AI developers’ responsibilities for ensuring attribution. These cases, albeit in the US, highlight the challenges AI poses to the right to attribution of authors whose works have been used as AI input.

EU copyright law under the InfoSoc and DSM Directives, however, regulates solely the economic rights of authors, leaving the protection of moral rights, including the right to attribution, up to the legislation of Member States. Consequently, even if AI developers can rely on TDM exceptions under the DSM Directive, this does not shield them from potential liability for infringing attribution rights. In the Pelham case, Advocate General Szpunar explicitly recognized this distinction in his opinion, referencing Deckmyn to assert that “[m]oral rights, particularly the right to the integrity of the work, may legitimately preclude the use of that work, even where that use is covered by an exception.” He reiterated this point in Spiegel Online, emphasizing that copyright exceptions under Article 5 of the InfoSoc Directive apply solely to economic rights and should not adversely affect any of the author’s moral rights, which include attribution. Arguably, the same logic applies to the exceptions to economic rights under the DSM Directive, meaning that the use of content as AI input under the TDM exceptions should respect the author’s right to attribution unless the applicable national exception explicitly covers moral rights. 

This consideration holds particular relevance for the lawful use of copyrighted works in AI training, given the longstanding tradition in European countries of strong protection for moral rights. In some Member States, such as France, moral rights hold near-sacred status, with courts historically enforcing robust protections for authors’ rights to integrity and attribution. Furthermore, the copyright laws of individual EU Member States are often ambiguous regarding the applicability of copyright exceptions to moral rights, leading to a lack of clarity as to the exceptions and limitations relevant to the right to attribution, including in the context of TDM. Consequently, even if AI developers lawfully rely on TDM exceptions, they may still face liability for moral rights infringement, as courts in certain jurisdictions may prioritize attribution and integrity regardless of TDM permissions. As attribution claims begin to surface in US cases involving generative AI, similar claims are likely to arise in Europe. National courts’ decisions on these matters could directly impact the lawful use of copyrighted works for AI training, potentially imposing additional requirements for attribution and integrity.

\subsubsection{Copyright and transparency rules under the AI Act}
The AI Act, in its final version adopted by the European Parliament on 13 March 2024, introduces two copyright-related provisions that apply to providers of general-purpose AI models (GPAIMs). These copyright provisions, found in Article 53(1)(c) and (d), aim to address the use of copyrighted content in AI training datasets, promoting transparency and compliance with existing copyright legislation. The first provision is directly related to the DSM Directive Article 4 TDM exception. According to Article 53(1)(c) AI Act, providers of GPAIMs must “put in place a policy to comply with Union copyright law, and in particular to identify and comply with, including through state of the art technologies, a reservation of rights expressed pursuant to Article 4(3) of Directive (EU) 2019/790”. The primary purpose of this obligation seems to be the practical implementation of the opt-out mechanism for TDM in the development of commercial generative AI models \cite{quintais2024generative}. While there are many open questions still surrounding the rightsholder reservation of rights, this new AI Act provision, as Quintais highlights, openly pushes for the standardization of such opt-outs. 

The second provision is the training content summary obligation set out in Art. 53(1)(d) AI Act, according to which providers of GPAIMs must “draw up and make publicly available a sufficiently detailed summary about the content used for training of the general purpose AI model”. It should be clarified that this obligation is a more general transparency requirement and is not limited to copyright-protected training data. The purpose of this provision is to facilitate the exercise and enforcement of the rights of all parties with legitimate interests, including copyright holders. In the context of copyright, this summary requirement can be useful to detect copyright infringements during the model development process, for example, if it lists a website that hosts a significant amount of infringing content, or the content rightsholders have opted out from its use for TDM \cite{peukert2024copyright}. 

The introduction of these copyright and transparency rules in the AI Act was met with criticism for multiple reasons, only some of which we shall highlight. First, by imposing copyright protections within a framework primarily aimed at mitigating AI risks to public health, safety, and fundamental rights, the AI Act integrates private economic concerns within a structure not designed for them. As Peukert notes, the EU legislator has in this way “merged two different types of laws”, complicating both compliance and enforcement. Additionally, Article 53 applies only to providers of GPAIMs, excluding other actors in the AI ecosystem, such as importers and deployers, who are otherwise subject to differentiated obligations in high-risk contexts under Chapter III of the AI Act. This limited scope may weaken enforcement, as it leaves the responsibility for copyright compliance solely with model providers and overlooks the role of downstream actors who deploy and use AI systems. Further clarity is expected from the AI Office and the upcoming AI Act Code of Practice, which may help address some compliance challenges. This guidance will be crucial in interpreting the new requirements, particularly what “sufficiently detailed” training data summaries mean, which currently remain vague and challenging for developers to implement. 

\subsubsection{Fair use and AI-generated content}
\label{subsubsec:fair-use-ai-gen-cont}
Example content summary: Legal framework, ongoing cases, DMCA protection against removal/modification of the Copyright Management Information; consequences for infringement; OpenAI cases; 

\paragraph{Introduction} 
Nowadays, legal complexities are surrounded by generative artificial intelligence (AI) and copyright, emphasizing challenges and litigation related to Copyright Management Information (CMI) under the Digital Millennium Copyright Act (DMCA). Generative AI systems, including models like ChatGPT and Stable Diffusion, are widely used across various industries, which raises intricate legal questions about copyright protection, especially when AI-generated outputs resemble copyrighted material. Central issues include the protection of CMI against unauthorized removal or modification, the liability implications for copyright infringement, and ongoing cases involving OpenAI and other AI developers \cite{garon2023practical, german2024copyright, volokh2023large, henderson2023foundation}.

\paragraph{Legal Framework}
The DMCA provides essential mechanisms to protect the CMI of digital works, ensuring that unauthorized alterations or removal do not obscure the ownership or creation details of copyrighted content. This framework is particularly relevant for generative AI models trained on extensive datasets that may contain copyrighted works. However, the absence of uniform international copyright norms leads to varied interpretations across jurisdictions, making compliance challenging for AI developers working with large, heterogeneous datasets \cite{lucchi2023chatgpt, sag2023copyright}.

\paragraph{Copyright Challenges in Generative AI} 
Generative AI models, which create content by deriving from large datasets that often include copyrighted texts, images, and other media, face significant copyright challenges. The use of this material without explicit authorization has led to numerous legal disputes. Notable cases, including Anderson v. Stability AI Ltd and Getty Images v. Stability AI, underscore the increased scrutiny of generative AI models. Courts frequently examine whether generative outputs constitute derivative works, applying fair use doctrine principles, especially when these outputs closely resemble the original content used for training \cite{german2024copyright, henderson2023foundation}.

\paragraph{Fair use and fair dealing}

In the United States, the fair use doctrine serves as a primary defence for many generative AI copyright cases. Courts assess factors such as the transformative nature of use, the amount of original material used, and the potential impact on the market for the original work. Generative AI systems often argue that their outputs are transformative; however, this argument is carefully scrutinized, particularly if the outputs closely mimic or directly reflect the copyrighted works in the training data. For instance, OpenAI’s ChatGPT has encountered defamation claims, raising questions about liability when AI-generated information contains false or defamatory statements about real individuals \cite{volokh2023large}.

In Europe, the concept of "fair use" as it exists in U.S. copyright law does not have a direct equivalent. Instead, European Union (EU) copyright law follows a "fair dealing" approach, which is more restrictive and allows specific exceptions and limitations on copyright use. These exceptions are narrowly defined and generally only apply in particular contexts such as research, criticism, news reporting, parody, and educational purposes.

In the EU, the InfoSoc and DSM Directivesshape the framework for copyright limitations. For example, Article 3 of the DSM Directive allows TDM for research organizations and cultural heritage institutions. This is significant for machine learning (ML) and AI, as it allows certain uses of copyrighted material without explicit permission from copyright holders, provided it is used solely for scientific research purposes. Article 4 extends TDM exceptions for any purpose, but copyright holders can opt out by explicitly prohibiting TDM in their terms of service, making it more limited than U.S. fair use \cite{german2024copyright, henderson2023foundation} (see also Section \ref{subsubsec:use-of-copyrighted-works-AI-EU}).

The EU's fair dealing framework means that generative AI developers need to carefully navigate copyright rules. Unlike in the U.S., where transformative use can justify broader use of copyrighted material, European jurisdictions often require express permission from copyright owners unless one of the specific exceptions applies. For instance, using large datasets containing copyrighted materials to train AI models without proper licensing may lead to infringement claims in Europe, even if the use is transformative or for the public good. This has prompted some countries, like the UK, to consider more AI-friendly copyright exceptions, though this is still an evolving area \cite{lucchi2023chatgpt}.

In practice, generative AI models trained in the EU must adhere closely to these narrow exceptions or obtain explicit permissions. The stricter interpretation of copyright in Europe means that developers cannot rely on broad "fair use" defences, which limits the scope for unlicensed AI training on copyrighted materials.

Recent legal actions underscore the tension between AI companies and content creators. In the UK, Mumsnet initiated legal proceedings against OpenAI, alleging that the company used data from its website without permission to train ChatGPT, thereby breaching copyright. Mumsnet claims that OpenAI scraped six billion words from its site without consent, posing an "existential threat" to websites like theirs[14]. Similarly, in the United States, authors including Sarah Silverman have filed lawsuits against AI developers like Meta, accusing them of using their copyrighted books without permission to train AI language models. These cases highlight the growing legal scrutiny of AI's reliance on copyrighted materials [15].

These developments indicate a global trend where content creators are increasingly challenging AI companies over the use of copyrighted material in training models. The outcomes of these cases could significantly impact how AI companies access and use data, potentially leading to more stringent regulations and the necessity for explicit licensing agreements.

The DMCA’s provisions against unauthorized CMI removal or alteration are crucial in AI applications where outputs might inadvertently strip metadata or alter original digital watermarks, potentially leading to liability. Violations of these provisions can result in both civil and criminal penalties, especially in cases where CMI manipulation is intentional. Liability may extend to AI models that systematically disregard the CMI attached to training data, creating calls for strict DMCA compliance to mitigate infringement risks \cite{german2024copyright, sag2023copyright}.
Ongoing legal cases underscore the generative AI industry’s exposure to copyright claims. Lawsuits against OpenAI, for example, include allegations that their models generate outputs resembling copyrighted material without proper attribution, which could impact the original creators’ markets. Plaintiffs argue that such outputs constitute unauthorized reproductions, advocating for “notice-and-takedown” mechanisms under the DMCA specifically adapted to AI-generated content as a means to manage infringement concerns \cite{volokh2023large, lucchi2023chatgpt}.

Industry experts recommend combining technical and legal mitigation strategies in response to these issues. AI developers are advised to implement technical solutions such as filtering outputs for close resemblances to training data or incorporating watermark detection to maintain CMI integrity. Additionally, some legal scholars propose AI-specific safe harbour provisions under the DMCA to protect models that demonstrate robust copyright compliance. Creating industry-wide guidelines could help ensure that AI-generated content adheres to copyright law, balancing technological innovation with the rights of original creators \cite{henderson2023foundation, sag2023copyright}.
As generative AI continues to integrate into commercial and creative industries, nuanced legal frameworks that address CMI protection, fair use, and liability are increasingly necessary. As legal precedents develop, finding a balance between copyright compliance and innovation will remain critical. Although the DMCA was initially designed for simpler digital environments, it now offers relevant tools to address the unique challenges posed by generative AI technologies and the extensive use of copyrighted data to train these models \cite{german2024copyright, sag2023copyright, garon2023practical}.

\section{From Legal Frameworks to Technological Solutions}
\label{sec:from-legal-to-technical}

\subsection{Generative Artificial Intelligence and LLMs} 
Generative Artificial Intelligence created a paradigm shift in AI development, enabling machines to create content that mimics human creativity across various mediums, including text, images, and video \cite{genAIgeneral}. State-of-the-art deep generative models like Generative Adversarial Networks (GANs), Variational Autoencoders (VAEs), Latent diffusion model (LDM), and transformers power many of today's cutting-edge applications. These models are widely used in creative industries, healthcare, and content generation, allowing for everything from image generation with tools like DALL-E to automating tasks in drug discovery. The underlying transformer architecture, the key to the progress in text generation, forms the basis for large language models (LLMs), which have set new benchmarks in natural language processing \cite{brown2020language, bommasani2021opportunities}. Large Language Models (LLMs) are models capable of understanding and generating human-like text. These models have billions of parameters, allowing them to perform complex tasks like translation, summarization, and legal document drafting. Their versatility is seen in real-world applications, such as customer service automation, personalized education, and legal compliance, where they reduce human intervention by performing repetitive tasks. The progress in GenAI and LLMs illustrates the growing potential of AI to not only replicate human reasoning and creativity but also to improve efficiency across multiple industries \cite{min2022recent,openai2023gpt4}.

\subsection{Attribution techniques}
\label{subsec:from-legal-to-technical}

\begin{table}[h]
\caption{Table gathering attribution techniques. }
\begin{center}
\begin{tabular}{ll}
\toprule
\textbf{Category} &
  \textbf{Technique} \\ \midrule
\textbf{Content Identification} &
  \begin{tabular}[c]{@{}l@{}}Watermarking\\ Fingerprinting\end{tabular} \\ \midrule
\textbf{Explainability methods} &
  \begin{tabular}[c]{@{}l@{}} Fine-tuning paradigm\\ Prompt paradigm\end{tabular} \\ \midrule
\textbf{Data traceability} &
  \begin{tabular}[c]{@{}l@{}}Data Lineage Tracking \end{tabular} \\ \midrule
\textbf{Security-Based Techniques} &
  \begin{tabular}[c]{@{}l@{}}Blockchain-Based Attribution\\ Digital signature\\ Zero-Knowledge Proofs\\ Synthetic Media Forensics\end{tabular} \\ \bottomrule
\end{tabular}
\label{tab:attribution_techniques}
\end{center}
\end{table}

In this section, we outline the methods that can be associated with the concept of attribution directly. Table \ref{tab:attribution_techniques} divides the attribution methodology into four areas: identification of generated content, model-based techniques, data traceability, and security-based techniques. Each of these techniques helps in identifying and tracking the content generated either from users or from AI models. While each of these methods presents distinct advantages, a comprehensive approach combining multiple techniques offers the most robust and practical solution to approximate reliable attribution of AI-generated content across diverse applications and use cases. Tracing methods, such as model activation traces and data lineage tracking, facilitate attribution by linking outputs to specific datasets or model configurations. These methods are valuable for auditing and ensuring compliance, particularly when combined with advanced explainability tools. However, the computational overhead and complexity can be limiting factors in real-time applications \cite{jiang2024watermarkbasedattributionaigeneratedcontent}  provides a systematic study on watermark-based, user-aware detection and attribution of AI-generated content, highlighting the importance of tracing techniques in attribution.
 
\subsection{Generated content identification}
The first category concerns the identification of content generated from AI models. In this category, the two main techniques, are watermarking and fingerprinting. Watermarking is based on inserting a marker into the content before it is released, while Fingerprinting is based on extracting a fingerprint from any content and comparing it to the fingerprint of another content. Watermarking involves embedding imperceptible yet traceable patterns in AI-generated outputs. These watermarks can take the form of specific word choices or token frequency patterns to ensure traceability without disrupting the user experience\cite{guo2024largelanguagemodelbased}. Research indicates its potential robustness against adversarial attempts, though challenges include resilience to content manipulation through paraphrasing or editing \cite{wen2023treeringwatermarksfingerprintsdiffusion, jiang2024watermarkbasedattributionaigeneratedcontent}. 

\paragraph{Watermarking} Watermarking involves embedding a traceable signature within AI-generated content to establish its origin. In the context of LLMs, watermarking can be implemented in several ways. One approach is to introduce imperceptible patterns or markers within the text generated by the model, such as specific word choices, sentence structures, or character frequencies. These markers would not be noticeable to human readers but could be detected using a specialized algorithm to verify the source of the content. For instance, a subtle adjustment of token probabilities or structured bias during text generation can serve as a digital watermark. The main advantage of this technique is that it provides a robust means to trace the source of the content, enhancing accountability and combating misinformation \cite{hagos2024recentadvancesgenerativeai} \cite{kalyan2023surveygpt3familylarge}. However, challenges include maintaining the invisibility and robustness of these watermarks. Adversarial actors could attempt to remove or alter watermarked content through paraphrasing or token manipulation, potentially undermining its traceability. Therefore, designing watermarks that are resilient to modification without significantly impacting content quality remains a key area of research. In their research, Zhang et al. \cite{zhang2023watermarks} show how strong watermarking is not possible for LLMs-generated output by introducing a novel watermark attack in which the attacker is not required to know the private key of the scheme or even which scheme is used.

\paragraph{Fingerprinting}
As defined by Wagner \cite{wagner1983fingerprinting} fingerprints are characteristics of an object that tend to distinguish it from other similar objects, and today fingerprinting can be used to identify LLM content. Pasquini et al. \cite{pasquini2024llmmap}, design a fingerprinting attack able to identify which model and version of an LLM is integrated into an application, revealing attack vectors that could compromise the system. Xu et al. \cite{xu2024instructional} present a pilot study on using lightweight instruction tuning as a form of LLM fingerprinting. The model publisher specifies a confidential private key and implants it as an instruction backdoor that causes the LLM to generate specific text when the key is present. Ribeiro et al. \cite{ribeiro2024leveraging} combine Fuzzy Fingerprints with LLMs and show that it is possible to obtain unique fingerprints for each author from the datasets.

\subsection{Explainability methods}
The second category is related to explainability methods for LLMs, which are classified as black-box models. Understanding, controlling, and applying these powerful models responsibly require explainable tools so-called Explainable Large Language Model (EXLLM). 

\subsubsection{Explainability in LLMs} 
In the review of Zhao et al. \cite{zhao2024explainability}, they divide explainability methods based on the type of training followed to build LLMs: traditional fine-tuning or prompting. 
\paragraph{Fine-tuning} Regarding traditional fine-tuning, they categorized techniques into local explanations and global explanations, proposing a division previously implemented with earlier Deep Learning (DL) models. Concerning the local explanations, feature-based explanations \cite{li2016understanding}, decomposition-based models \cite{montavon2018methods} and gradient-based \cite{enguehard2023sequential} could be highlighted besides the ones based on surrogate models. Surrogate models, which rely on simpler models to explain the output of a more complex model, have several paradigms of explanation, mainly as Shapley Additive Explanations (SHAP) \cite{lundberg2017unified} and Local Interpretable Model-Agnostic Explanations (LIME)  \cite{ribeiro2016should,nikolic2024experimental}. Explainability techniques in the fine-tuning realms include decomposition-based models aimed at breaking the input contribution  \cite{montavon2018methods}. A particular type of these explanation methods can be associated with the concept of attribution, the so-called data influence category. It characterizes the influence of individual training samples by measuring the loss effect on test points \cite{yeh2018representer}, and represents a direct connection between the output model and the training data. Focusing on global explanations, the aim is to describe the inner workings of the model, without focusing on the single explanations. Probing-based explanations are one of the global methods that describe the LLMs acquired knowledge during the training process \cite{raffel2020exploring, marvin2018targeted} and examine how the model encodes knowledge across layers by training simple classifiers (probes).

\paragraph{Prompting paradigm}
As highlighted by Zhao et al., the more LLMs scale, the more they show linguistic capabilities. However, this brings the necessity to change the methods used to explain their outputs, as previous methods become ineffective. Therefore, new methodologies are settled as explaining In-Context Learning (ICL) \cite{li2023survey}, and Chain of Thought (CoT) Prompting \cite{madaan2022text}, which respectively focus on how ICL and CoT influence model behaviour, and representation engineering approaches, which focus on structure and the characteristics of the representation space to capture emergent representations and high-level cognitive phenomena \cite{marks2023geometry}. Another typology of explanation is represented by the paradigm of Assistant Model Explanation, which includes explaining the role of fine-tuning \cite{zhou2024lima}, explaining hallucination \cite{dziri2022origin}, and uncertainty quantification \cite{xiong2023can}. 

\subsection{Data traceability}
Another field which can contribute to the attribution process is data traceability, which refers to the process of tracking the origin, the process and the destination of data while it migrates between different sources. 

\paragraph{Data Lineage Tracking}
Karkovskova et al. \cite{karkovskova2021design} introduce the design and practical application of the Business data lineage model as a method of metadata management to manage and model data processing. The model consists of three layers that provide a conceptual, logical, and physical view of the business data lineage. As it is nontrivial to discover the data flow/movement from its source to destination, such that monitoring different transformations and hops on its way in an enterprise environment, Tang et al. \cite{tang2019sac} build Spark-Atlas-Connector (short as SAC), a new system to track data lineage in a distributed computation platform. SAC provides a visual representation of data lineage to track data from its origin to downstream and is deployed in a distributed production environment to demonstrate its efficiency and scalability.

\subsection{Security-based techniques}

The last category relates to security-based methods. The most important categories in the field of attribution are blockchain methods, digital signatures, zero-knowledge proofs, and synthetic media forensics. Security-based approaches approaches offer tamper-proof attribution mechanisms. By integrating different approaches security-based methods ensure integrity and traceability while safeguarding user privacy. Recent applications in secure advertising and healthcare demonstrate their practical viability. Liu et al. \cite{liu2023blockchainempoweredlifecyclemanagementaigenerated} present a blockchain-empowered framework to manage the lifecycle of edge AI-generated content products, highlighting the role of cryptographic signatures in attribution. 

\paragraph{Blockchain attribution}
As in the era of digital marketing it is even more important to attribute the right amount of work to each published, Liu et al. \cite{liu2023blockchain}, approach this problem by combining blockchain with advertising attribution to solve the problem of attribution, allowing publishers and advisers to store data in real-time, guaranteeing good results as because blockchain is decentralized, tamper-proof, and traceable. Chakraborty et al. \cite{chakraborty2019value} propose an approach to attribute the value brought by new members introduced in the blockchain and gained by the members participating in it. Their tracking method, other than the fine-grained provenance permits them to share the value back with the contributing organizations in a fair and decentralized manner.

\paragraph{Cryptographic signatures} Cryptographic signatures are mathematical schemes used to ensure the authenticity of messages. Recent research employed cryptographic signatures to secure user prompts \cite{gupta2024enhancing}. Their method ensures that system interactions are possible only to authenticated users. Research watermarking, metadata tagging, model output tracing, and cryptographic signatures. The manuscript by \cite{sgantzos2023triple} proposes using Triple-Entry Accounting (TEA), to address concerns related to the ethical use of Large Language Models (LLMs) and intellectual property in the context of dataset collection. By applying TEA to a publicly accessible distributed ledger, the proposal aims to control LLM queries, prevent malicious actions, and protect intellectual property rights. The use of digital signatures enhances the integrity and verifiability of transactions, making it easier to audit and share records, thus mitigating ethical concerns in AI and data usage.

\paragraph{Zero-knowledge proofs for AI}
Zero-knowledge proofs consist of a protocol in which one party convince another party that some given statement is true, without conveying to the verifier any information beyond the mere fact of that statement's truth. This type of protocol is starting to be applied also to applications involving AI and LLMs. Cai et al. \cite{cai2023hcpp} propose a Healthcare Chatbot-based Privacy Preserving (HCPP) Framework that adopts a data-oriented approach to reduce the excessive disclosure of personal information. HCPP consists of the Healthcare Chatbot-based Minimized Personal Information (HCMPI) method and the Healthcare Chatbot-based Zero Knowledge Proof (HCZKP) method. HCMPI leverages LLMs to minimize the acquisition of unnecessary personal health information without significantly affecting healthcare services. HCZKP further encrypts a part of the minimized information, making the data available but invisible. Wang et al. \cite{wang2024efficient} propose a blockchain-based distributed framework for Large Language Models (LLMs) to address data security concerns. In their system, the outputs of the LLMs are stored and verified on the blockchain, ensuring integrity, transparency, and traceability. A multiparty signature authentication mechanism is used to ensure consensus among stakeholders before publication. To enhance security, they introduce a threshold elliptic curve digital signature algorithm that works in environments with three or more participants. The approach incorporates discrete logarithmic zero-knowledge proofs (ZKPs) and Feldman verifiable secret sharing. 

\paragraph{Synthetic Media Forensics}
In the realm of attribution, a related topic is the detection of deepfakes, as the detection allows one to discern between subjects with the willingness to generate the content and content manipulated without the willingness of the protagonist. Jia et al. \cite{jia2024can}, investigate the capacity of ChatGPT to detect deepfakes, finding that LLMs can detect generated deepfakes even though they are not directly trained for that operation. Another relevant research area in the realm of media forensics is the analysis of fake news. In their literature review, Kuntur et al. \cite{kuntur2024under} shed light on the field of fake news detection by providing a brief historical overview, defining fake news, reviewing detection methods used before the advent of large language models, and discussing the strengths and weaknesses of these models in an increasingly complex landscape. Furthermore, it will emphasize the importance of using multimodal datasets in the effort to detect fake news.

\subsection{Discussion of Applicability and Organisational Engagement of Technical Methods for Attribution} This section reflects on the existing technical methods for attribution and presents the engagement of those by AI organisations. In more depth, leading AI organisations, such as Microsoft, Google, NVIDIA, Meta, Adobe, IBM, and Palantir, have started adopting the aforementioned techniques. This is in response to regulations on AI technology, and specifically AI-generated content, as well as governmental and political pressures. 

In more detail, Google DeepMind developed a tool, called SynthID, to watermark generative AI outputs  \cite{Smith_2024}. Specifically, following up on last year's watermarking mechanism for images, this year it introduced a tool for watermarking AI-generated video that is publicly available on Hugging Face. This is available on the Gemini app and online chatbots, while it is compatible with multiple other platforms. Additionally, other big AI companies, including Microsoft, Meta, and Amazon have been committed to include watermarks in their AI-generated content \cite{watermarking}. 

Furthermore, Adobe, Arm, Microsoft, Intel, and OpenAI support metadata tagging to maintain transparency in AI outputs, especially in content shared across social media and creative platforms \cite{metadata}. On model output tracing techniques, NVIDIA, Meta, Microsoft, Google and Palantir suggested that they will engage with techniques for tracing outputs back to models \cite{tracing}. This movement was triggered, inter alia, by the White House AI plan \cite{whitehouse}, which aimed to manage data authenticity and track model usage, enhancing accountability for generative outputs. Following the White House AI plan, Microsoft and IBM are exploring cryptographic signatures and are committed to developing secure signatures under the voluntary AI standards supported by the U.S. government, ensuring that AI-generated content maintains a verifiable trail to its source  \cite{tracing}. 

Overall, those commitments do not guarantee the adoption of those techniques throughout all the operations of those organisations, they are a first step towards maintaining transparency and accountability. Nevertheless, those companies cannot be fully trusted to regulate themselves, and therefore both governments should develop novel human-driven laws for AI-generated content, as well as humans should develop a critical eye and demand transparency and credibility. 

\section{Legal cases as Use Cases for Attribution Methods}
\label{sec:legal-cases}

This section outlines how the technical solutions identified under Section \ref{subsec:from-legal-to-technical} can be applied to facilitate copyright enforcement and licensing. We present two use cases that show how attribution technology can help copyright holders understand when their content is being used for AI training and whether this is done according to the applicable legislation or contractual terms. Use case one explores how attribution methods can be used to track and enforce licensing schemes and TDM ‘opt-outs’. Use case two focuses on the use of such techniques as evidence in copyright infringement cases. 

\subsection{Attribution Methods, Licensing and Opting-out}

The growing copyright litigation in the US, coupled with the remaining uncertainties regarding the practical application of the DSM Directive’s TDM opt-out provision and the new AI Act copyright rules in the EU, has led to a rise in licensing agreements between AI model providers and rightsholders. Such agreements are concluded usually between technology companies and either controllers of aggregate copyright works (news publishers \cite{niemanLab}, stock image companies \cite{ventureBeat}) or collective management organizations (CMOs) \cite{alcssurvey2024}. 

According to the European Copyright Society “In effect, the AI training space is already moving to licensing as a default” \cite{dusollier2025copyright}. In this shifting context, the practical role of attribution techniques is most prominent in enabling the identification of works used in AI training datasets and their status (opted-out, licensed, or free-to-use). Attribution methods can, therefore, be invaluable for monitoring and enforcing licensing terms throughout the AI model’s life cycle. 

When an AI developer enters a licensing agreement (e.g., obtaining permission to train on a stock photo library or a news archive) both parties have an interest in ensuring the data is used according to the agreed terms. Digital watermarking, fingerprinting, and even blockchain-based attribution can create an auditable trail of content usage. 

A digital watermark embedded in licensed training images, for example, would persist (imperceptibly) in the AI model’s parameters or outputs, allowing the licensor to later verify whether their images were indeed used and whether the model’s outputs appropriately reflect the licensed status. If the model outputs an image or a fragment that carries the unique marker, it confirms the use of the licensed dataset and can even identify the source. Conversely, if the model outputs licensed content without the expected watermark or in ways not permitted (for example, generating an image that is a near copy of a licensed input when only transformative uses were allowed), it would flag a potential breach of the license. 

In the text domain, model fingerprinting techniques can help track the usage of licensed literary works by recognizing distinctive patterns or tokens originating from those works in the AI’s output. Blockchain-based attribution, on the other hand, could play a role by recording each transaction or use of a licensed work on a tamper-proof ledger. For instance, every time a licensed news article is fed into a model or generates an output, a blockchain entry could log the event (including which content ID was used and under which license), creating an immutable record. These attribution technologies can thus act as compliance monitors, giving licensors confidence that their content is handled per the contract and alerting them to misuse.

Several real-world licensing models illustrate how attribution technology can support new business models based on AI licensing. 

\subsubsection{The Shutterstock Case}

Shutterstock’s AI licensing program is a prominent example. Shutterstock not only licensed millions of images to OpenAI (for training image generation models) but also implemented a system to compensate artists whose works are included. Contributors are paid 20\% of the licensing revenue and can opt out of having their content used for AI training \cite{ventureBeat}. This arrangement can be achieved only with robust tracking: Shutterstock must know which specific images (by which contributors) were used by the AI. This can be done by attaching metadata or identifiers to each image in the licensed dataset so that downstream usage can be tallied for compensation. Similarly, Getty Images has partnered with NVIDIA to develop a “commercially safe” generative AI model trained exclusively on Getty’s licensed library of photos \cite{gettyimagesai}. By using only images that carry Getty’s licensing and watermark information (or other rights metadata), the model reduces the risk of untraceable or infringing outputs. Getty’s initiative presumably leverages content ID technology to ensure that only permitted images feed into the model, and any output can be traced back to authorized training data.

\subsection{Attribution Methods and Copyright Infringement}

Attribution techniques can be used in practice not only for licensing schemes but also in copyright infringement litigation involving AI. When allegations arise that a generative AI system has unlawfully used copyright-protected works, digital attribution can provide the forensic evidence needed to support or refute those claims. A key legal question in these cases is often whether the AI model’s output is derived from specific copyrighted inputs (in the EU, relevant is also whether these inputs have been opted-out of AI training). Technical attribution methods make tracing to the original copyrighted material possible. For instance, if a dataset included images with a subtle digital watermark uniquely identifying the rightsholder, and elements of that watermark are detected in the AI’s output, this can be considered compelling evidence that the image was used in training without permission. 

In addition, as pointed out under Section \ref{subsec:legal-status-ai-gen-works}, the WCT and WPPT oblige signatory countries (including EU member states and the US) to protect electronic rights management information. That means a party who knowingly strips digital identifiers from a work (for example, deleting authorship metadata before feeding the work into an AI) can be held liable for that act independent of the direct infringement. In the context of AI, if a developer were to ignore a creator’s TDM reservation (opt-out) expressed via a metadata tag or watermark, this could be seen as akin to removing CMI. This would mean the developer not only infringed the reproduction right by using the work without authorisation but also violated laws protecting the integrity of copyright management information. In court, evidence of such removal or disregard of a technical attribution marker can undercut a defence that the AI company’s use was innocent or protected by an exception.

\subsubsection{The Getty Case}

Several cases already exemplify how technical attribution evidence can decisively influence the outcome of copyright disputes involving AI. The first, famously discussed in the Getty Images cases against Stability AI \cite{iaia2024elephant}, is watermark image detection. 

In these cases, Getty argues that the presence of their digital watermark in Stable Diffusion’s output provides clear, almost irrefutable evidence of unauthorized training in Getty’s library. This argument implies that no amount of abstraction by the AI model can erase the fact that specific protected images were used; the watermark’s appearance is a forensic ‘smoking gun’. Moreover, if it’s shown that Stability AI developers attempted to remove or obscure watermarks from training images, Getty could argue that there was a knowing removal of copyright management information. Thus, watermarking technology can be used both to detect the infringement and form the basis of a legal violation itself.

\subsubsection{Doe vs GitHub case}
Another example is the potential use of digital fingerprints to catch unlicensed training data. For example, an author can place a hidden fingerprint in the text of her works (e.g., a particular sequence of punctuation or a unique string of words at set intervals recognizable to algorithms). If an AI model later produces a paragraph in the style of that author with the exact hidden sequence placed as a fingerprint, this would strongly indicate that the model was trained on the author’s novel despite, say, the author’s public TDM opt-out notice on her website. A real-world parallel can be seen in court cases regarding code-generation AI models, in which the plaintiffs’ claim is based on fingerprint-related arguments. 

In Doe v. GitHub, for example, developers sued GitHub’s Copilot for regurgitating licensed code with unique comments or whitespace patterns intact, effectively fingerprinting the source repository \cite{farcon2024attribution}. As fingerprinting and other attribution techniques evolve, we can expect their wider use for demonstrating copyright infringement. For example, invisible noise patterns embedded in training images that survive generative transformations, or cryptographic hashes of text fragments that can be recognized in model weights. All these attribution techniques can serve the same purpose - providing courts and rightsholders with a clear chain linking AI outputs to specific inputs, thereby enabling enforcement.

\section{Discussion}
\label{sec:discussion}
As LLMs are revolutionizing virtually every field, from research to defence, from education to artistic output, they rely on an enormous quantity of data to be trained. However, is often the case that the data employed to train LLM is used without the authorization of the creators, thus posing an enormous barrier to the possibility of recognizing the work of the people that generated the data. 

In this manuscript, we analyze tools available to achieve \emph{attribution}, which is the practice of tracing and connecting the content generated by LLMs to the data used to train them, thus attributing the content generated by LLM to the people who created the training data. 

This part of the manuscript is dedicated to analyzing the copyright laws and the legal framework at the base of the right of content creators to recognise that their content was used to train LLMs. Moreover, the first part of the manuscript highlights the contemporary legal status of AI-generated works. From a legal perspective, copyright and intellectual property laws differ significantly across jurisdictions, complicating the creation of universal standards for LLM attribution. 

In the second part of the manuscript, we categorize the techniques that allow the implementation of the laws that regulate attribution. We focus on the techniques that allow us to trace the content from the moment in which it is created, from the moment in which the data is used to train the model \cite{karkovskova2021design,tang2019sac}. Subsequently, we focus on the techniques that allow us to connect the generated content to the training data \cite{zhao2024explainability, li2016understanding,montavon2018methods, enguehard2023sequential,lundberg2017unified, ribeiro2016should,nikolic2024experimental}. 

However, the effectiveness of these tools depends on their widespread adoption, and willingness to regulate and enforce their use. Due to this, in the third and last part of the manuscript, we use three use cases (the Shutterstock case, The Getty Case, Doe vs GitHub case) to depict how the fusion of technological and legislative tools allows us to reach attribution, by providing a foundation for accountability, intellectual property protection, and transparency. 

At the same time, we highlight how bridging the gap between legal frameworks and technological solutions for attribution mechanisms in large language models (LLMs) is indeed a complex endeavour. While there are tools and methods from both legislative and technological perspectives that offer potential solutions, the integration of these domains faces significant hurdles. The rapid pace of AI advancements often outstrips the ability of current legal structures to adapt, leaving gaps in enforcement and creating ambiguity regarding ownership, accountability, and responsibility. This, in turn, poses challenges in establishing clear guidelines for attributing content or actions to specific entities or algorithms.

Technologically, while attribution mechanisms are crucial for ensuring transparency and fairness, they must navigate a host of issues related to privacy, security, and bias. AI systems, particularly LLMs, operate in intricate ways that make it difficult to trace their outputs to specific inputs, and even harder to establish clear lines of responsibility. 

Despite these challenges, the need for effective attribution in AI systems is critical, not only for ensuring the ethical use of these technologies but also for fostering trust in AI systems and their outcomes. As such, it is imperative that both lawmakers and technologists collaborate closely to develop frameworks and tools that are flexible, scalable, and adaptable to future advancements in AI. Only through such a multidisciplinary approach can we hope to achieve a balanced and effective system of attribution in the rapidly evolving landscape of AI.

\section{Conclusions and Future Work}
\label{sec:conclusion}

In our review, we showed that the convergence of law and technology is essential in establishing effective LLM attribution. Through the use cases, we showed how the convergence of intents between law and technology can allow the implementation of attribution, with positive consequences for both the creators of the content used to train AI models and for the subjects responsible for their implementation and training. 

While legal frameworks must adapt to the evolving AI landscape, incorporating provisions that recognize AI-generated content while ensuring accountability and fairness, technological solutions must be refined to offer scalable and secure attribution methods. By integrating these domains, we can foster a balanced approach that upholds ethical standards, protects intellectual property, and promotes responsible AI usage.

Future research should focus on developing standardized attribution protocols that are legally enforceable and technologically robust. Collaborative efforts between policymakers, law experts, and technologists will be instrumental in shaping regulations which adapt to technological advances. As LLMs evolve, a proactive and interdisciplinary approach will be key to addressing attribution challenges unifying the legal, research, and technical powers under one dimensioned framework, ensuring the responsible deployment of AI-generated content. 

\section{Authors' Contributions}

Legal background and legal use cases (Sections \ref{sec:legal-background}, \ref{sec:legal-cases}): SL (lead), NV, SG, SR. 

LLMs background and technological solutions (Section \ref{sec:from-legal-to-technical}): EM (lead), AM, MM, and RH.  

Visualization and formatting: EM.

Supervision: RH.


\end{document}